\documentclass{aa} 
\usepackage{graphics} 
\begin{document} 
\baselineskip=1.2em 
\thesaurus{02.01.1; 02.19.1; 02.20.1; 09.03.2; 09.04.1; 13.07.1} 
\title{Proton and electron acceleration through magnetic turbulence in 
relativistic outflows}\author{R. Schlickeiser \inst{1} 
\and C. D. Dermer \inst{2}} 
\date{Received 16 February 2000 / Accepted 15 May 2000} 
\institute{$^1$ Institut f\"ur Theoretische Physik, Lehrstuhl IV:  
Weltraum- und Astrophysik, Ruhr-Universit\"at Bochum,  
D-44780 Bochum, Germany; r.schlickeiser@tp4.ruhr-uni-bochum.de\\ 
$^2$E. 
O. Hulburt Center for Space Research, Code 7653, Naval Research Laboratory, 
Washington, DC 20375-5352 
USA; dermer@osse.nrl.navy.mil} 
\offprints{R. Schlickeiser} 
\authorrunning{Schlickeiser \& Dermer} 
\titlerunning{Particle acceleration by magnetic turbulence} 
\maketitle 
 
\begin{abstract} 
 
Low frequency electromagnetic turbulence is generated in relativistically 
outflowing plasma that sweeps up particles from the surrounding environment. 
Electrons are energized by stochastic gyroresonant acceleration with the 
turbulence produced by the isotropization of captured protons and charged 
dust. Protons are accelerated stochastically by the dust induced turbulence.  
Analytical solutions for the proton and electron energy distributions are 
obtained and used to calculate broadband synchrotron emission.  The 
solutions are compared with generic spectral behavior of blazars  
and gamma-ray bursts. 
Dust captured by a blast wave can generate turbulence that could accelerate 
protons to very high energies. 
 
\end{abstract} 
 
\section{Introduction} 
 
Measurements of radio and optical polarization of blazar jet sources 
indicate that nonthermal lepton synchrotron emission is the dominant 
radiation mechanism at these frequencies (Begelman et al.\ \cite{begelman}). 
The radio through X-ray afterglow  and the prompt soft-gamma ray emission 
observed (Costa et al.\  \cite{costa}, van Paradijs et al.\ 
\cite{vanparadijs}, Djorgovski et al.\ \cite{djorgovski}, Frail et al.\ 
\cite{frail}) from gamma-ray bursts (GRBs) is attributed to the same process 
(Tavani \cite{tavani}, Paczy\'nski \& Rhoads \cite{paczynski}, M\'esz\'aros 
\& Rees \cite{mr93}). In the standard blast-wave physics that has been 
developed (M\'esz\'aros \& Rees \cite{meszaros97}; Vietri \cite{vietri97}, 
Waxman \cite{waxman97})  to explain the long wavelength afterglow emission 
from GRBs, particles captured from the external medium energize the blast 
wave and cause blast wave deceleration. When a blast wave sweeps up material 
from the surrounding medium, the free energy of the captured particles 
initially resides in the more massive protons, nucleons, or charged grains 
and dust, if the latter can survive capture by a relativistic wind, as 
assumed here.  The less massive nonthermal electrons and positrons emit, 
however, most of the radiant energy.  In order that high radiative 
efficiencies are possible, it is therefore required that there is efficient 
transfer of energy from the swept-up massive particles to electrons, or to very 
high-energy protons emitting synchrotron radiation or photomeson secondaries. 
 
Major uncertainties in modeling relativistic outflows involve magnetic field 
generation and the mechanism for dissipating and transferring the free 
energy of protons, ions and dust particles to lighter particles 
(M\'esz\'aros, Rees \& Papathanassiou \cite{meszaros93}, Chiang \& Dermer 
\cite{chiang}). Nonthermal electrons and protons may both undergo 
first-order Fermi shock acceleration in relativistic blast waves to form 
power-law distributions reaching to very high energies, and the origin of 
ultra-high energy cosmic rays (UHECRs) has been attributed to shock 
acceleration of protons in the GRB blast waves  (Vietri \cite{vietri}, 
Waxman \cite{waxman}).  This mechanism has recently been called into 
question (Gallant \& Achterberg \cite{gallant}). Following the first shock 
crossing by a particle, subsequent cycles do not permit large gains of 
energy because the particle is captured by the shock before it has been 
scattered through a large angle. A stochastic mechanism for particle 
energization in relativistic shocks avoids these difficulties, but requires 
strong magnetic fields to retain UHECRs within the shell (Waxman 
\cite{waxman}). However, the detailed physics of stochastic gyroresonant 
acceleration was not considered. 
 
In this paper, we show that the magnetic turbulence generated by the capture 
of heavy charged particles can accelerate lighter charged  
particles to high energies through 
stochastic gyroresonant acceleration. This could account for the hard 
spectra observed from GRBs and blazars during their flaring states. The 
acceleration of particles to high energies could produce UHECRs through 
gyroresonant acceleration in relativistic outflows. The question of the 
survivability of dust against sublimation by the intense radiation fields 
of the GRB is addressed in the Appendix. 
 
\section{Particle energization in relativistic outflows} 
\subsection{The relativistic pick-up model} 
A two-stream instability is formed when a blast wave with an entrained 
magnetic field encounters a medium of density $n_i^*$. According to the 
recent work of Pohl \& Schlickeiser (\cite{pohl}, hereafter referred to as 
PS\cite{pohl}), this causes the captured protons and electrons to rapidly 
isotropize in the blast wave plasma. The isotropization process operates on 
the comoving time scale (eq. (74) of PS\cite{pohl})  $t_{\rm iso} ({\rm s}) 
\cong 0.7 n_{b,8}^{1/2}/\Gamma_{300}n_i^*$, where $\Gamma = 300 
\Gamma_{300}$ is the bulk Lorentz factor of the blast wave and $n_{b,8}$ 
denotes the comoving blastwave density in units of $10^8$ protons cm$^{-3}$. 
The blast wave particle 
density in the comoving frame is given by $n_b ({\rm cm}^{-3})\sim$ $ 
E/(\Gamma_0 m_pc^2 4\pi r^2 \Delta r)$$\sim 10^8 E_{54}/r_{16}^3$, and is 
implied by depositing an initial energy $10^{54}E_{54}$ ergs/(4$\pi$ sr) in 
a blastwave shell with comoving radius $\Delta r \sim r/\Gamma_0$, where 
$\Gamma_0$ is the initial blastwave Lorentz factor. 
Alfvenic magnetic turbulence is generated by the incoming protons and 
electrons with an energy density (eq. (53) of PS\cite{pohl}) $\Delta U_A 
\cong \beta_A m_p c^2 n_i^* \Gamma \sqrt{\Gamma^2-1}$, where $\beta_A c$ is 
the Alfven speed.  The total energy density of the swept-up material in the 
comoving frame is $\Delta U_{i} = m_p c^2 n_i^* \Gamma \sqrt{\Gamma^2-1}$ 
(Blandford \&  McKee \cite{bm76}). Thus the fraction of incoming energy 
converted to transverse plasma wave turbulence in the process of 
isotropizing the particles is $\sim \beta_A$. 
 
PS\cite{pohl} calculated the emitted radiation in the thick-target limit 
when the captured, isotropized protons undergo nuclear inelastic collisions 
with thermal particles in the blastwave.  Although this process involves the 
fraction $\sim (1-\beta_A)$ of the incoming energy retained in the protons, 
it operates on the comoving pion production time scale $t_{pp}({\rm s}) \sim 
1.4\times 10^7/n_{b,8}$, where $n_{b,8}$ denotes the comoving blastwave 
density in units of $10^8$ protons cm$^{-3}$. If the acceleration rate is 
sufficiently rapid, pion production could produce low-level prompt 
and extended emission in GRBs or blazars if $n_b$ remains at 
the estimated level. Moreover, luminous radiation from pion-decay electrons 
and positrons is only produced if $t_{pp} $ is much shorter than the 
diffusive proton escape time scale from the blast wave, given by $t_{\rm 
esc} \cong (\Delta r)^2/\kappa = r^2/(\Gamma_0 ct_{\rm iso})^2$, noting  
that the spatial diffusion coefficient $\kappa \cong c^2 t_{\rm iso}$. This 
requires that $n_{b,8}^{1/2}n_i^*r_{16}^2>>15\Gamma _{300}$, which holds 
well in jets but is only marginally satisfied in GRB blast waves. 
 
\subsection{The role of captured charged dust} 
Here we extend the work of PS\cite{pohl} by considering the process of 
channeling the plasma turbulence energy into the energy of the swept-up 
primary electrons through gyroresonant interactions.  As demonstrated in 
that paper, incoming charged particles with mass $m_j$, charge $Q_j e$, and 
Lorentz factor $\Gamma$ generate Alfvenic turbulence with parallel 
wavenumbers $|k| > R_j^{-1}$, where the Larmor radius $R_j = m_j c^2 
\sqrt{\Gamma^2 -1}/(Q_j|e| B_0)$ and $B_0$ is the  
entrained magnetic field in the frame of the 
blastwave. For simplicity, we assume that the magnetic field is parallel to 
the blast wave velocity, and generalize the calculation of PS\cite{pohl} by 
including  
the effects of negatively 
charged dust. According to eqs. (48)-(50) of PS\cite{pohl}, we  
obtain the power spectrum of 
backward ($-$) and forward (+) propagating Alfven waves given by $I_-(k) 
\cong |Z(k)|$ and $I_+(k) \cong I_0^2(k)/|Z(k)|$, where $I_0(k)$ is the 
initial turbulence spectrum before isotropization, and  
\begin{displaymath} 
Z(k) = -\;{1\over 2}\;{\omega_{p,i}\over c} {n_p\over n_b} {B^2\over k^2} 
\end{displaymath} 
\begin{equation} 
\hskip1.0cm\times \sum_{j=e,p,d} {Q_j n_j \over n_p} [1-(R_j k)^{-1}]\; 
H[\;|k|-R_j^{-1}]\;. 
\label{wavespectrum} 
\end{equation} 
 
The term $\omega_{p,i} ({\rm s}^{-1})= 1.3\times 10^3\sqrt{n_b}$ is the 
proton plasma frequency and $H[x]$ denotes the Heaviside step function. 
This 
expression assumes $I_0(k)  \ll |Z(k)|$ and makes use of the 
charge-neutrality condition $Q_d n_d + n_e = n_p = \Gamma n_i^*$, where we 
assume that all dust particles have the same charge. 
A number of processes influence the charging of dust grains in a 
proton-electron-plasma. In the absence of any other process due to the higher 
electron mobility an initially uncharged dust particle will absorb more 
electrons than protons and thus attain a negative potential of order 
$-2.5k_BT_e/e$ (Spitzer 1968). However, a number of competing processes 
like neutral-atom collisions and ion impacts on the grain surface 
lead to electron loss from the dust grains. Also, a significant ultraviolet 
photon flux can charge the grain by the photoelectric effect to a positive 
potential. All these processes, together with poorly known grain properties, 
make an exact determination of the grain charge impossible (see e.g. 
McKee et al. (1977) for the variety of grain properties resulting from  
model calculations under various assumptions). Here we follow the argumentation 
of Ellison, Drury \& Meyer (1997). If the grain potential is $\phi $, then 
the charge on a spherical (of radius $a$) grain is  
of order $q=eQ_d\simeq 4\pi \epsilon _0a\phi $, so that  
$Q_d\simeq 700 (a/(10^{-7}{\rm m}))(\phi /10 {\rm V})$. 
The number of atoms in the grain will be of order $[a/(10^{-10} {\rm m})]^3$, 
so that with the mean atomic weight $\mu $ of the grain atoms 
the entire grain atomic weight is $A_G=\mu [a/(10^{-10} {\rm m})]^3$. 
This implies a dust mass of $m_d=\mu m_p[a/(10^{-10} {\rm m})]^3= 
2\cdot 10^{10}m_p\, a_{-7}^3$ if we adopt $\mu =20$ for silicate grains. 
Combined with the charge estimate this yields for the  
charge-weighted dust/proton mass ratio  
$Y_d= m_d/(Q_d m_p)=3\cdot 10^7\, a_{-7}^2$.

 It is important to note 
that the densities of the incoming protons, electrons and dust particles 
are much smaller than the density of the blast wave plasma $n_b^*$, so 
that the properties of the low-frequency Alfven waves, carried by the blast 
wave plasma, are not modified by the captured particles. Mathematically,  
this corresponds to the statement that the real part of the Alfvenic 
dispersion 
relation is solely determined by the blast wave electrons and protons. 
However, the free energy of beams of incoming protons, electrons and dust 
particles gives rise to a non-zero imaginary part of the Alfvenic dispersion 
relation leading to a positive growth rate of backward moving (in the 
blast wave plasma) Alfven waves. Moreover, for relativistic beams  
(Miller \cite{miller}) the growth rate of longitudinal electrostatic waves 
is much smaller than the growth rate of transverse Alfvenic waves, 
so that the free energy is dissipated by isotropising the incoming beam 
(i.e. capture) rather than heating the blast wave plasma by plateauing. 
 
Two comments on the equilibrium wave spectrum (1) are appropriate: 
 
\noindent (1) According to the quasilinear relaxation theory the equilibrium  
Alfven wave spectrum given in Eq. (1) is achieved formally after  
infinitely long time $t_{\rm ql}\to \infty $ by transferring the free energy in the 
initial particle-beam-distributions into plasma waves. Numerical simulations 
of the electrostatic beam instability (see e.g. Grognard 1975) indicate  
that this asymptotic equilibrium distribution is established after 
$t_{\rm ql}\simeq 100t_{\rm iso}$. 
 
\noindent (2) The wave spectrum in Eq. (1) is calculated from quasilinear  
wave kinetic equations that only take into account the wave growth 
from the unstable particle beams but neglect wave-wave interactions. Since 
the latter scale with the total magnetic field fluctuation energy  
density $(\delta B)^2$, this is  
justified as long as the total wave energy density is small to the initial 
energy density of the beam particles. Because according to  
Eqs. (53) and (64) of 
PS\cite{pohl} $(\delta B)^2/(4\pi w_0)=\beta _A<<1$ this is indeed fulfilled.

The turbulence spectrum (1) leads to the quasilinear isotropization  time 
scale  
$t_{\rm iso}$ given above. This time scale and $t_{\rm ql}=100 t_{\rm iso}$  
are much shorter than the 
blastwave crossing time $t_{lc}({\rm s})\sim \Delta r/c \sim 1000 
r_{16}/\Gamma_{300}$.  Fig. 1 shows the resulting turbulence 
spectrum of 
backward propagating, left-handed circularly polarized Alfven waves for a 
charge-weighted dust/proton density ratio $\epsilon = Q_d n_d/n_p = 0.01$, 
and a charge-weighted dust/proton mass ratio $Y_d$ $ = m_d/(Q_d m_p)$ $= 
10^6$. The 
relative contributions of dust to the turbulence spectrum (1) is determined 
primarily by the parameters $\epsilon$ and $Y_d$. Most of the magnetic 
turbulence energy density resides at low wave numbers and is provided by the 
charged dust for this value of $\epsilon$. 
 
\begin{figure} 
\vskip-0.8in 
\resizebox{\hsize}{!}{\includegraphics{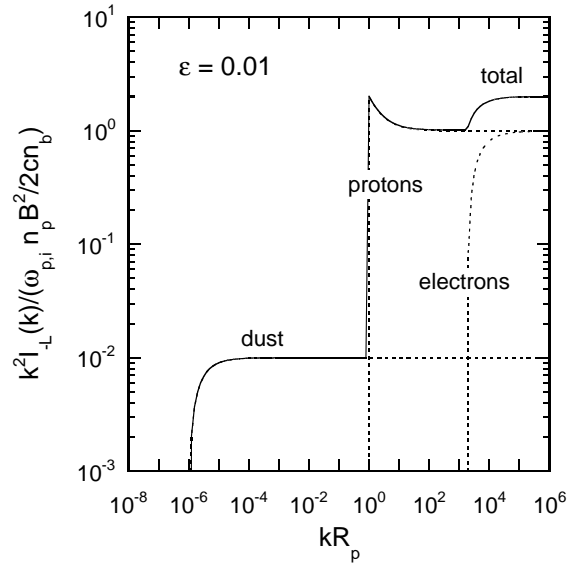}} 
\caption{Intensity $I_-(k)$ of backward propagating Alfven waves at wave 
vector $k$ generated  by a two-stream instability when charged dust, 
protons, and electrons are captured by a relativistic blast wave. 
Parameters: $\epsilon = Q_dn_d/n_p = 0.1$ and $Y_d = m_d/(Q_dm_p) = 10^6$. } 
\label{fig1} 
\end{figure} 
 
The total dust mass $M_d = m_d n_d = Y_d\epsilon M_p$, where the total 
proton mass $M_p= m_p n_p$. Thus the total mass in dust for the calculation  
in Fig. 1 is $\sim 10^4$ times the mass in the ionized gas, as 
might be found in the weakly ionized, low temperature  
star-forming environments that harbor GRB sources. By contrast, the relative 
mass in dust to that in protons would be much less in the enviroments of 
active galactic nuclei, where the intense UV and X-ray radiation field would 
photo-disintegrate the dust. 
 
\subsection{Stochastic gyroresonant acceleration of electrons and protons} 
The primary electrons and protons may increase their energy by stochastic 
gyroresonant acceleration with the turbulence spectrum given by 
Eq.({\ref{wavespectrum}).  Since a fraction $\beta _A$ of the incoming  
free energy is converted into transverse plasma wave turbulence, these 
particles can tap at most a fraction $\beta_A$ of the swept-up energy via 
this process. Efficient energy transfer therefore requires that 
$\beta_A\gg 0.01$. The e-folding time scale for stochastic gyroresonant 
acceleration is $t_{\rm acc} \sim \beta_A^{-2} t_{\rm iso}$, whereas the 
diffusive escape time $t_{\rm esc} \cong R^2/\kappa = t_{lc}^2/t_{iso}$ 
(Barbosa \cite{barbosa}, Schlickeiser \cite{schlickeiser89}). 
Because the turbulence spectrum in 
Eq.({\ref{wavespectrum}) is $\propto k^{-2}$, the acceleration, escape, and 
isotropization time scales are independent of particle energy.  The comoving 
deceleration time scale $t_{\rm dec}$(s)$ = 1.3\times 10^4 
(E_{54}/n_i^*\Gamma_{300}^5)^{1/3}$ (M\'esz\'aros \& Rees \cite{mr93}). 
Provided that 
 $\beta_A \ge 0.03n_{b,8}^{1/4}(\Gamma_{300}^2/E_{54}n_i^{*2})^{1/3}$, the 
acceleration time scale is much shorter than the time scale for evolutionary 
changes of the blast wave due to bulk deceleration. In this case, a 
steady-state calculation of the equilibrium particle energy spectrum is 
justified, which results from the balance of gyroresonant acceleration, 
diffusive escape, deceleration and radiative losses. 
 
The steady-state kinetic equation for the phase-space density $f_e$ of 
relativistic electrons is (Schlickeiser \cite{schlickeiser84}) 
\begin{equation} 
{1\over\gamma^2}\;{d\over d\gamma}[{\gamma^4\over t_{\rm a}} {df_e\over 
d\gamma} + ({\gamma ^3\over t_{\rm dec}}+{\gamma^4\over t_{\rm syn}}) f_e]\; 
 - \; {f_e\over t_{\rm E}} = 
-S_e\delta(\gamma-\Gamma)\;  
\label{kinetice} 
\end{equation} 
where $S_e = \sqrt{\Gamma^2 -1} n_e^*c/(4\pi \Gamma r)$ and $A$ 
is the area of the blast wave that is effective in sweeping up material.  
 
If no dust is present, then the electrons can only be accelerated to $\gamma 
< 
\Gamma R_p/R_e = \Gamma m_p /m_e$. In this range, $t_a = t_{\rm acc}$ and 
$t_E= t_{\rm esc}$. If dust is present, then electrons can be accelerated 
throughout the range $\Gamma m_p/m_e \leq  \gamma < \gamma_{\rm e,max} = 
\Gamma m_d/(m_e Q_d)$, but the acceleration and escape times are modified 
according to the relations $t_a = t_{\rm acc}/\epsilon$ and $t_E= \epsilon 
t_{\rm esc}$. The radiative loss time scale is $t_{\rm syn} ({\rm s}) = 6\pi 
m_e c/(\sigma_T B_0^2) = 7.7\times 
10^8 B_0^{-2}$.  
  
Likewise, the steady-state kinetic equation for the phase-space density 
$f_p$ of relativistic protons is 
\begin{displaymath} 
{1\over\gamma^2}\;{d\over d\gamma}[{\epsilon \gamma^4\over t_{\rm acc}} 
{df_p\over d\gamma} + \gamma^3({1\over t_{pp}}+{1\over t_{\rm dec}}) f_p]\;  
\end{displaymath} 
\begin{equation} 
\hskip1.0cm - \; {f_p\over \epsilon 
t_{\rm esc}} 
= -S_p\delta(\gamma-\Gamma)\; , 
\label{kineticp} 
\end{equation} 
where $S_p = \sqrt{\Gamma^2 -1} n_i^*c/(4\pi \Gamma r)$. 
Eq.(\ref{kineticp}) holds in the presence of dust when $\gamma < 
\gamma_{\rm p,max} $ $= \Gamma R_d /R_p = $ $\Gamma m_d /(m_p Q_d)$ $= 
\Gamma Y_d$, and assumes that effects of photopion production 
do not limit the acceleration of protons to the highest energies.  
Evidently, the parameter $Y_d$ and the bulk Lorentz factor $\Gamma $ determine 
the maximum proton Lorentz factor $\gamma _{\rm p,max}$. 
If no dust is present, then protons are not significantly accelerated by 
gyroresonant processes unless 
long wavelength MHD turbulence is generated through processes not treated 
here.  
 
The exact solution to Eq.(\ref{kinetice}) for electrons accelerated by 
turbulence with a spectrum given by Eq.(\ref{wavespectrum}) is  
\begin{displaymath} 
f_e(\Gamma < \gamma < \gamma_{\rm e,max}) = t_a S_e  
{{\tilde \Gamma}[\mu -(3-a)/2]\over 
{\tilde \Gamma}[1+2\mu]}\;  
\end{displaymath} 
\begin{displaymath} 
\hskip1.0cm {\Gamma^{\mu+(a+1)/2}\gamma^{\mu-(a+3)/2}\over 
\gamma_s^{2\mu}} 
\exp (-\gamma/\gamma_s) \times 
\end{displaymath} 
\begin{equation} 
\hskip0.2cm M(\mu - {3-a\over 2}, 1+2\mu,{\Gamma\over \gamma_s})U(\mu - 
{3-a\over 2}, 
1+2\mu,{\gamma\over \gamma_s}) 
\label{esolt} 
\end{equation} 
(Schlickeiser \cite{schlickeiser84}),  
where $\mu = \sqrt{\lambda + (3-a)^2/4}$, $\lambda = t_a/t_E$,  
$a=t_a/t_{\rm dec}$, $\tilde \Gamma (a_0)$ denotes the gamma function, and 
$M(a_1,a_2,a_3)$ and $U(a_1,a_2,a_3)$ are the Kummer functions of the first 
and second kind, respectively.  The value $\gamma_s = t_{\rm syn}/t_{\rm 
acc} \cong 370 \Gamma_{300}n_i^*/n_{b,8}^{3/2}$ is the electron Lorentz 
factor where the electron synchrotron loss time scale equals the 
acceleration time scale. In the regime $\gamma \gg \Gamma$, an accurate 
approximation is obtained by taking the asymptotic expansion of the function 
$U$ for small arguments, resulting in  
\begin{displaymath} 
f_e(\Gamma < \gamma < \gamma_{\rm e,max}) \cong 
\end{displaymath} 
\begin{equation} 
 \hskip1.0cm {t_a S_e\over 2 \mu} \; 
\Gamma^{\mu+ (1+a)/2}\;\gamma^{-(3+a)/2-\mu}\;\exp (-\gamma/\gamma_s)\;. 
\label{esoltapprox} 
\end{equation} 
 
The exact solution to Eq.(\ref{kineticp}) for protons accelerated by 
turbulence with a spectrum given by Eq.(\ref{wavespectrum}) is 
\begin{displaymath} 
f_p(\Gamma < \gamma < \gamma_{\rm p,max}) =  
\end{displaymath} 
\begin{equation} 
\hskip1.0cm {t_{\rm acc} S_p\over 2\epsilon 
\mu_p} \; \Gamma^{\mu_p+(a_p+1)/2}\;\gamma^{-(a_p+3)/2-\mu_p}\;, 
\label{psolt} 
\end{equation} 
where $\mu_p = \sqrt{\lambda_p+(3-a_p)^2/4}$, $\lambda_p = t_{\rm 
acc}/\epsilon^2 t_{\rm esc}$, and  
$a_p = t_{\rm acc}[t_{\rm pp}^{-1}+t_{\rm dec}^{-1}]/\epsilon $. 
 
\begin{figure} 
\vskip-0.8in 
\resizebox{\hsize}{!}{\includegraphics{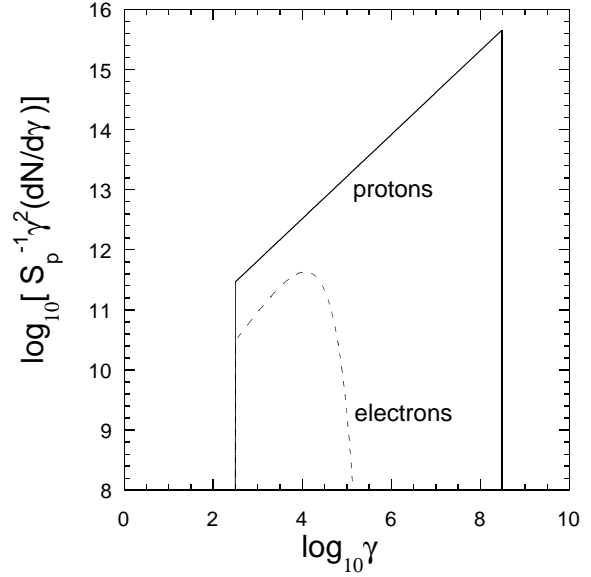}} 
\caption{Steady-state distribution of electrons and protons  
for $\gamma \ge \Gamma =300$  
resulting from 
stochastic gyroresonant acceleration with waves generated in a relativistic 
blast wave for parameters given in text. } 
\label{fig2} 
\end{figure} 
 
In Fig. 2, we show the resulting proton and electron spectra, 
choosing 
$\Gamma_{300} = 1$, $\beta_A = 0.03$, $r_{16} = 0.1$, $n_{b,8} = 0.1$, 
$\epsilon = 0.1$, $Y_d = 10^6$, and $n_p^* = 1$ cm$^{-3}$. As can be seen, 
the electron spectrum $dN_e/d\gamma  \propto \gamma^{-1}\exp 
(-\gamma/11700)$ is very flat below the synchrotron cutoff, whereas the 
proton spectrum displays a spectrum $dN_p/d\gamma  \propto \gamma^{-1.36}$ 
below the maximum Lorentz factor $\gamma_{\rm p, max} = 3\times 10^8$. For 
these parameters, we find comoving acceleration times of $\sim 330 
(\beta_A/0.03)^{-2}$ s for electrons, and $\sim 10$ times larger for the 
protons. The choosen value for $\beta _A$ implies a magnetic field 
strength of 14 Gauss. Larger values of $\beta_A$ will shorten the 
acceleration time scales 
and improve the accelerated particle efficiency.  
Requiring that the gyroradii of the swept-up dust particles $R_d=Y_dR_p$ 
in a magnetic field of strength $14 B_{14}$ Gauss is smaller than the comoving 
shell radius $10^{14}r_{16}\Gamma _{100}^{-1}$ cm  
limits the charge-weighted dust/proton mass ratio to values 
smaller than $Y_d<5\cdot 10^6\, r_{16}\, B_{14}\, \Gamma _{100}^{-2}$, 
and consequently the maximum proton Lorentz factor in the comoving 
blast wave frame is  
$\gamma _{p,\rm{max}}=\Gamma Y_d<5\cdot 10^8\, r_{16}\, B_{14}\,  
\Gamma _{100}^{-1}$.} 
 
\begin{figure} 
\vskip-0.8in 
\resizebox{\hsize}{!}{\includegraphics{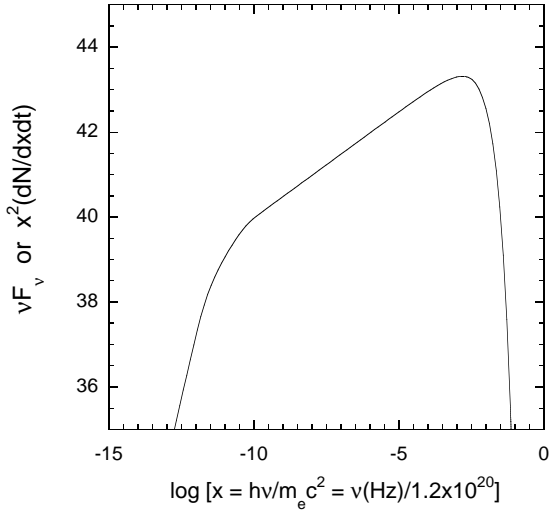}} 
\caption{Synchrotron emission spectrum in a $\nu F_\nu$ representation for 
the electron distribution given in Fig. 2.} 
\label{fig3} 
\end{figure} 
 
Fig.\ 3 shows the resulting synchrotron emission spectrum produced by the 
electron 
distribution function (5).  For the parameters used 
here, the 
synchrotron spectrum is very hard.  Particle spectral indices 
$dN/d\gamma\propto \gamma^{-p}$ with $p \geq 1$ can be produced through this 
process, leading to synchrotron spectra $F_\nu \propto \nu^{-\alpha}$ with 
$\alpha = -1/3$, corresponding to the index of the elementary synchrotron 
emissivity formula for monoenergetic electrons.  
In standard shock acceleration theory (Blandford \&  
Eichler 1987), particles are accelerated with $p \geq 2$, leading to values 
of 
$\alpha \geq 1/2$.   
Harder synchrotron spectra can 
however be produced through nonlinear effects in first-order Fermi 
acceleration, or by introducing a low-energy cutoff in the electron 
distribution function. 
Synchrotron spectra from flaring blazar sources (Catanese et al.\ 
\cite{catanese}, Cohen et al. \cite{cohen},  
Pian et al.\ \cite{pian}) and GRBs 
display very hard spectra requiring nonthermal lepton spectra with $p < 2$, 
consistent with the stochastic Fermi acceleration mechanism proposed here. 
 
The hardening above the ankle of the cosmic ray spectrum and beyond the  
ZGK cutoff could be related 
to a hard UHECR component formed through stochastic acceleration in GRB 
blast waves (Waxman \cite{waxman}, Takeda et al.\ \cite{takeda}). 
Because turbulence is generated in the process of capturing and isotropizing 
particles from the external environoment, a wave spectrum that can yield 
very high energy electrons and protons through gyroresonant acceleration 
must be present in the relativistic blast waves. If dust survives capture by 
the magnetized plasma wind, then long wavelength turbulence can be generated 
that could accelerate protons to ultra-high energies on time scales shorter 
than the duration over which the blast wave decelerates to nonrelativistic 
speeds.  Gyroresonant acceleration of particles by MHD turbulence provides a 
mechanism 
to transfer energy from heavier to lighter species, and overcomes 
difficulties associated with accelerating 
UHECRs in relativistic blast waves through the first-order Fermi mechanism. 
 
\section {Summary and conclusions} 
 
We have shown that efficient and rapid stochastic gyroresonant acceleration 
of  
electrons and protons occurs when interstellar protons and dust are captured 
by a relativistic outflow. The MHD plasma wave turbulence, created within 
the  
blast wave plasma by the penetrating  
protons and charged dust particles, is transferred from the 
heavier to the lighter particles through the gyroresonant 
acceleration process. This mechanism is particularly attractive 
for the energization of nonthermal electrons in GRB sources, as 
radiation modelling often requires approximate energy equipartition between 
electrons and protons (e.g., Katz \cite{katz}, Beloborodov \& Demianski 
\cite{beloborodov},  
Smolsky \& Usov \cite{smolsky}, Chiang \& Dermer \cite{chiang}).  
Capture by and isotropization within a blast wave, as discussed by Pohl \&  
Schlickeiser (\cite{pohl}), provides primary electrons with an energy 1836 
times  
smaller than that of the protons. As shown here, the subsequent acceleration 
of electrons 
in the proton-induced turbulence leads to approximate energy equipartition 
of these nonthermal electrons. Moreover, if in addition charged dust 
particles 
penetrate the blast wave plasma, we find that further acceleration of the 
particles  
is possible to an energy determined by the 
radiative loss processes of the electrons and protons. For some parameter 
regimes, 
this could accelerate particles to $\sim 10^{20}$ eV and account for UHECR 
production. 
 
By solving the appropriate kinetic equations, we calculated the resulting  
particle energy spectra of the accelerated protons and electrons, and the 
observable synchrotron radiation spectrum of the accelerated electrons.  
The latter peaks at hard X-ray energies and is very flat, in agreement with 
observed photon spectra of GRBs and some blazars during their flaring 
states. 
 
A crucial issue, particularly for the stochastic acceleration of protons, 
is the penetration of dust. While large amounts of dust are present 
in the environments of star-forming regions 
in galaxies, it is also necessary that the 
dust not be destroyed during capture by the blast wave or by the intense 
radiation associated with captured nonthermal particles in the blast wave. 
We address some aspects of this issue in Appendix A. 
 
\begin{acknowledgements} 
RS gratefully acknowledges partial support of his work by the 
Deutsche Forschungsgemeinschaft through Sonderforschungsbereich 191 and 
by the Verbundforschung, grant DESY-05AG9PCA. The work of CD 
is supported by the Office of Naval Research and NASA grant DPR S-13756G. 
\end{acknowledgements}

\section {Appendix A: Survivability of dust in GRB radiation field} 
 
A major uncertainty in this study is whether the dust can survive the  
intense GRB radiation field by sublimation before the blast wave  
arrives and sweeps up the dust particle. In the inhomogeneous  
environment surrounding a massive star/GRB progenitor (Dermer \&  
B\"ottcher \cite{db00}), it is probable that the blast wave will impact a  
dense dusty cloud after passing through a nearly evacuated region.  In  
that case, there will be little radiation in advance of the blast wave,  
because the blast wave will only start to be energized and radiate  
after it encounters the cloud. Under such circumstances, the  
survivability of a large fraction of the dust to sublimation is  
assured. 
 
In the ideal case where a GRB blast wave passes through a uniform  
density medium before intercepting dust particles, we outline a  
calculation of dust sublimation and identify parameter sets that are  
compatible with dust survivability.  We follow the treatment of Waxman  
\& Draine (\cite{wd99}) for dust sublimation. Although reverse shock  
emission could be important for dust sublimation, we neglect it here.   
It has only been detected from GRB 990123 and can be strongly  
suppressed if, for example, the blast wave shell is thin and the  
reverse shock traverses the shell before becoming relativistic (Sari et  
al.\ \cite{snp96}). 
 
In this case the optical-UV emission irradiating the dust, which is the  
photon energy range most important for dust sublimation (Waxman \&  
Draine \cite{wd99}), can be described by the $L_\nu$ = 
$L_{\nu,{\max}} (\nu/\nu_0)^{1/3}$ portion of the synchrotron  
emissivity spectrum  where the quantity $L_{\nu,{\max}}$ = $4\pi  
n_i^* x^3 m_ec^2 \sigma_{\rm T} \Gamma B_0/9e$ (Sari et al.\  
\cite{spn98}), where $e$ is the proton charge. Using standard blast-wave 
physics, $B_0$(G)  
$= 0.388 \Gamma\sqrt{n_i^* e_B}$, where $e_B < 1 $ is the magnetic-field  
parameter. To give good fits to the prompt phase of GRBs, it is  
necessary that the electrons are in the weakly cooled regime (Chiang \&  
Dermer \cite{chiang}, Dermer et al.\ \cite{dbc00}). For a weakly cooled  
spectrum, $\nu_0 = \nu_m = \Gamma\gamma_m^2 eB/2\pi m_e c$, where  
$\gamma_m = e_e\Gamma (p_e-2)m_p/[m_e(p_e-1)]$ and $e_e  >\sim 0.1$.  
Taking the electron injection index $p_e \approx 2.2$ - 3.0 = 2.5, as  
implied by burst spectroscopy in the prompt and afterglow phase,  
we find $\nu_m$(Hz) =$4.1\times 10^{11}e_e^2 \Gamma^4 \sqrt{n_i^* e_B}$  
(see Sari et al.\ \cite{spn98}, Dermer et al.\ \cite{dbc00} for more  
details). The luminosity (ergs s$^{-1}$) between the 1 eV $= 2.41\times  
10^{14}$ Hz and 7.5 eV band is 
\begin{equation} 
L_{1-7.5} \cong 1.28\times 10^{-5} n_i^{*4/3} (e_B/e_e^2)^{1/3} x^3  
\Gamma^{2/3} \cong 
\label{Lopt} 
\end{equation} 
\begin{displaymath} 
2.0\times 10^{46} ({n_i^*\over 100})^{4/3} ({e_B\over 10^{-4}})^{1/3}  
({e_e\over 0.5})^{-2/3}({x\over 10^{16}{\rm cm}})^3({\Gamma \over  
300})^{2/3}\;,\end{displaymath} 
which also defines our standard parameter set. 
 
The relationship of $\Gamma$ and $x$ to observer time $t$ for an  
adiabatic blast wave decelerating in a uniform surrounding medium is 
\begin{equation} 
{\rm~~~~~}{\Gamma\over \Gamma_0} = {1\over  
\sqrt{1+(4\tau)^{3/4}}}\;,{\rm~and~} {x\over x_d} \cong {\tau\over  
1+4^{-1/4}\tau^{3/4}}\; , 
\label{kin} 
\end{equation} 
where the deceleration radius $x_d = (3E_0/8\pi\Gamma_0^2m_pc^2  
n_i^*)^{1/3}$ $\cong$ $2.1\times 10^{16}(E_{54}/\Gamma_{300}^2  
n_2)^{1/3}$ cm, the dimensionless time $\tau = t/t_d = ct\Gamma_0^2 c/x_d$,  
and the apparent isotropic explosion energy is $10^{54}E_{54}$ ergs. 
 
For the sublimation of dust, we consider the total energy change due to  
(1) heating by the GRB in the 1 - 7.5 eV band, (2) thermal reradiation  
by dust, and (3) the energy carried away by the sublimed particles. In  
this approximation, the heating  rate $dT/dt$ due to dust energization  
is given through 
\begin{displaymath} 
[4\pi a^3 k_{\rm B}({\rho\over m})]^{-1}{dT\over dt} = {dE\over dt}  
= {L_{1-7.5}(t)a^2(t)\over 4\pi r^2}\;- 
\end{displaymath} 
\begin{equation} 
4\pi a^2(t) \sigma_{\rm SB}T^4 - 12\pi k_{\rm B}T({\rho\over m})a^2(t)  
\cdot {da\over dt}\; . 
\label{dTdt} 
\end{equation} 
The rate of shrinkage of dust grain radius $a(t)$ with time $t$, is given  
by 
\begin{equation} 
{da\over dt} = - ({m\over \rho})^{1/3} \nu_0 \exp(-b/k_{\rm B}T)\;, 
\label{dadt} 
\end{equation} 
(Waxman \& Draine \cite{wd99}), where $\rho/m = 10^{23}$ cm$^{-3}$,  
$\nu_0 = 1\times 10^{15}$ s$^{-1}$, and $b/k_{\rm B} = 7\times  
10^4$ K for refractory grains. 
 
We look for solutions of Eqs. (\ref{Lopt})-  
(\ref{dadt}) where the dust grain is closer than the blast wave to the  
GRB source at the moment the dust becomes completely sublimed. There  
are no interesting solutions for our standard parameters; in this case,  
dust at $10^{16}$ cm becomes sublimed when the blast wave has only  
reached $x = 1.6\times 10^{15}$ cm. It is easy, however, to find  
solutions for slightly less energetic GRBs in more tenuous  
environments. For example, the blast wave will reach dust particles  
before they sublime if the dust resides within $x  \cong 9\times  
10^{15}$ cm when $E_0 = 10^{52}$ ergs and $n_i^* = 1$ cm$^{-3}$. 
 These values of the energy and density are not too different than the 
values  
found when modeling the afterglow of GRB 980425 (Wijers \& Galama  
\cite{wg99}). Thus we find that dust will survive destruction by  
sublimation in some, though not all, GRB environments if the  
surrounding medium is uniform. A clumpy surrounding medium improves the  
chances of dust survival.

{}

\end{document}